\documentclass{article}

\usepackage{graphicx,cite}

\begin{document}

\title{The minimal length uncertainty and the quantum model for the stock market}

\author{    Pouria Pedram\thanks{p.pedram@srbiau.ac.ir}\\
            {\small Department of Physics, Science and Research Branch,
            Islamic Azad University, Tehran, Iran}
            }

\date{\today}
\maketitle \baselineskip 24pt

\begin{abstract}
We generalize the recently proposed quantum model for the stock
market by Zhang and Huang to make it consistent with the discrete
nature of the stock price. In this formalism, the price of the stock
and its trend satisfy the generalized uncertainty relation and the
corresponding generalized Hamiltonian contains an additional term
proportional to the fourth power of the trend. We study a driven
infinite quantum well where information as the external field
periodically fluctuates and show that the presence of the minimal
trading value of stocks results in a positive shift in the
characteristic frequencies of the quantum system. The connection
between the information frequency and the transition probabilities
is discussed finally.
\end{abstract}


\section{Introduction}
Econophysics as an interdisciplinary research field was started in
mid 1990s by physicists who are interested to apply theories and
models originally developed in physics for solving the complex
problems appeared in economics, specially in financial markets
\cite{1}. Because of the stochastic nature of the financial markets,
the majority of tools for market analysis such as stochastic
processes and nonlinear dynamics have their roots in statistical
physics. Besides statistical physics, other branches of physics and
mathematics have a major role in the development of econophysics.
The sophisticated tools developed in quantum mechanics such as
perturbation theory, path integral (Feynman–-Kac) methods, random
matrix and the spin-glass theories are shown to be useful for option
pricing and portfolio optimization. Among theoretical physics,
quantum field theory has a special role to reveal the intricacies of
nature from quantum electrodynamics to critical phenomena. For
instance, it can be used to model portfolios as a financial field
and describes the change of financial markets via path integrals and
differential manifolds \cite{4,5}.

The application of quantum mechanics to financial markets has
attracted much attention in recent years to model the finance
behavior with the laws of quantum mechanics and it is becoming now a
rather established fact \cite{9,10,11,12,13,14,15}. For instance,
Schaden, contrary to stochastic descriptions, used the quantum
theory to model secondary financial markets to show the importance
of trading in determining the value of an asset \cite{6}. He
considered securities and cash held by investors as the wave
function to construct the Hilbert space of the stock market. Another
useful application of quantum theory to trading strategies is
quantum game theory which is the generalization of classical game
theory to the quantum domain \cite{7,8}. This theory is primarily
based on quantum cryptography and contains superimposed initial wave
functions, quantum entanglement of initial wave functions, and
superposition of strategies in addition to its classical
counterpart.

At this point, it is worth explaining why quantum mechanics is
essential to study the behavior of the stock market. Classical
mechanics which is described by Newton's law of motion is
deterministic in the sense that it exactly predicts the position of
a particle at each instant of time. This is similar to the evolution
of a stock price with zero volatility ($\sigma=0$) that results in a
deterministic evolution of the stock price. However, in  the context
of quantum mechanics, the evolution of the position of the particle
has a probabilistic interpretation which is similar to the evolution
of a stock price with a non-zero volatility ($\sigma\ne0$) \cite{5}.
Note that there is a close connection between the
Black-Scholes-Merton (BSM) equation \cite{Black,Merton} and the
Schr\"odinger equation: The position of a quantum particle is a
random variable in quantum mechanics, and similarly, the price of a
security is a random variable in finance. Also, the Schr\"odinger
equation admits a complex wave function, whereas the BSM equation is
a real partial differential equation which can be considered as the
Schr\"odinger equation for imaginary time. Haven showed that BSM
equation is a special case of the Schr\"odinger equation where
markets are assumed to be efficient \cite{Haven}. Indeed, various
mathematical structures of quantum theory such as probability
theory, state space, operators, Hamiltonians, commutation relations,
path integrals, quantized fields, fermions and bosons have natural
and useful applications in finance. In the language of Schaden,
``The evolution into a superposition of financial states and their
measurement by transaction is my understanding of quantum finance''
\cite{Schaden}.

Recently, Zhang and Huang have proposed a new quantum financial
model in econophysics and defined wave functions and operators of
the stock market to construct the Schr\"odinger equation for
studying the dynamics of the stock price \cite{Zhang}. They solved
the corresponding partial differential equation of a given
Hamiltonian to find a quantitative description for the volatility of
the Chinese stock market. In their formalism, the wave function
$\psi(\wp,t)$ is considered as the price distribution, where $\wp$
denotes the stock price and $t$ is the time. There, the stock price
is approximately considered as a continuous-variable. However, the
stock price is actually a discrete variable and admits a non-zero
minimal price length $\left(\Delta \wp\right)_{min}\ne 0$ which
depends on the stock market's local currency. In this paper, we
incorporate the fact of discrete nature of the stock price with the
quantum description of the stock market. We show that the
uncertainty relation between the price and its trend and the form of
the Hamiltonian should be modified to make the quantum formulation
consistent with discrete property of the stock price. Note that,
Bagarello has also tried to present quantum financial models which
describe quantities which assume discrete values
\cite{11,12,13,14,15}. However, Bagarello's approach is mainly based
on the Heisenberg approach rather than on the Schr\"odinger
equation.

\section{The Quantum Model}
Before applying quantum theory to finance we need to identify the
macro-scale and micro-scale objects of the stock market. Since the
stock index is based on the share prices of many representative
stocks, it is meaningful to consider the stock index as a macro
system and take every stock as micro-scale object \cite{Zhang}. Note
that each stock is always traded at a certain price which shows the
particle behavior. Also, the stock price always fluctuates in the
market which is the wave property. Therefore, because of this
wave--particle duality, we can consider the micro-scale stock as a
quantum system. Now we can construct the quantum model for the stock
market based on the postulates of quantum mechanics.

First, we introduce the wave function $\psi(\wp_0,t)$ as the vector
in the Hilbert space which describes the state of the quantum
system. More precisely, $\psi(\wp_0,t)$ is the state vector
$|\psi,t\rangle$ in price representation i.e. $\psi(\wp_0,t)\equiv
\langle \wp_0|\psi,t\rangle$. Also, we take the modulus square of
the wave function as the price distribution and demand that the
superposition principle of quantum mechanics also holds
\begin{eqnarray}
|\psi\rangle=\sum_n c_n |\phi_n\rangle,
\end{eqnarray}
where $|\phi_n\rangle$ is the possible orthonormal basis states of
the stock system and $c_n =\langle \phi_n|\psi\rangle$. Therefore,
the state of the stock price before trading should be a
superposition of its various possible states with different prices
so-called a ``wave packet''. We can consider a trading process, buy
or sell at some price, as a physical observation or measurement. So
the trading process projects the state of the stock to one of the
possible states with a definite price where $|c_n|^2$ denotes its
probability. In other words, we can interpret $|\psi(\wp_0,t)|^2$ as
the probability density of the stock price versus time, namely,
\begin{eqnarray}
P(t)=\int_a^b|\psi(\wp_0,t)|^2d\wp_0,
\end{eqnarray}
which shows the probability of the stock price between $a$ and $b$
at time $t$.

At this point, we introduce the Hermitian operators correspond to
the financial observables in the Hilbert space. The price operator
$\hat{\wp_0}$ measures the price of a state vector similar to the
position operator $\hat{x}$ in quantum mechanics. Similarly, we
define $T_0=m_0 \displaystyle \frac{d\wp_0}{dt}$ as the rate of
price change which is related to the trend of the price in the stock
market\footnote{Note that this definition is only valid at the
continuous price limit. As we shall show in the next section, when
we consider the minimal price length, $T$ cannot be written simply
as the derivative of the price with respect to time. However, it can
be expressed in terms of such operators (\ref{x0p01}).} where $m_0$
is the mass of stock. As indicated by Baaquie, a comparison between
the Black-Scholes Hamiltonian and the Schr\"odinger equation shows
that the volatility is the analog of the inverse of mass i.e.
$m_0=\sigma^{-2}$ (see Sec.~4.4 of Ref.~\cite{5}). Also, the
intensity of the price movement can be considered as the energy of
the stock or the associated Hermitian Hamiltonian which results in
the following Schr\"odinger equation
\begin{eqnarray}
\hat{H}\psi(\wp_0,t)=i \frac{\partial}{\partial t}\psi(\wp_0,t),
\end{eqnarray}
where the Hamiltonian is a function of price, trend, and time and
generates the temporal evolution of the quantum system.

The trade of a stock can be considered as the basic process that
measures its momentary price. This measurement can only be performed
by changing the owner of the stock which represents the Copenhagen
interpretation of a quantum system \cite{6}. Therefore, a
measurement may change the outcome of subsequent measurements so
that it cannot be described by ordinary probability theory. Indeed,
we can never simultaneously know both the ownership of a stock and
its price. The stock price can only be determined at the time of
sale when it is between traders. Moreover, the owners decide to sell
or buy the stock at higher or lower prices that determines the trend
of the stock price. So, in the quantum domain, the stock price and
stock trend operators satisfy the following uncertainty relation
\begin{eqnarray}
\Delta\wp_0\Delta T_0\geq 1/2,
\end{eqnarray}
where $T_0=-i\partial/\partial\wp_0$. However, as we show in the
next section, this relation should be modified when we take into
account the discrete nature of the stock price.

\section{The Generalized Uncertainty Principle}
According to the above uncertainty relation, in principle, we can
separately measure the price and the trend with arbitrary precision.
However, since the price is a discrete variable there is a genuine
lower bound on the uncertainty of its measurement. Thus, the
ordinary uncertainty principle should be modified to so-called
generalized uncertainty principle (GUP). Here we consider a GUP
which results in a minimum price uncertainty
\begin{eqnarray}\label{gup}
\Delta \wp \Delta T \geq \frac{1}{2} \left( 1 +\beta_0 (\Delta T)^2
+\zeta \right),
\end{eqnarray}
where $\beta_0$ and $\zeta$ are positive constants which depend on
the expectation value of the price and the trend operators. In
ordinary quantum mechanics $\Delta \wp$ can be made arbitrarily
small as $\Delta T$ grows correspondingly. However, this is no
longer the case if the above relation holds. For instance, if
$\Delta \wp$ decreases and $\Delta T$ increases, the new term
$\beta_0 (\Delta T)^2$ will eventually grow faster than the
left-hand side and $\Delta \wp$ cannot be made arbitrarily small.
Now the boundary of the allowed region in $\Delta \wp\Delta T$ plane
is given by
\begin{eqnarray}
\Delta T =\frac{\Delta
\wp}{\beta_0}\pm\frac{1}{\beta_0}\sqrt{(\Delta
\wp)^2-(1+\zeta)\beta_0},
\end{eqnarray}
which yields the following minimal price uncertainty
\begin{eqnarray}\label{zeta}
(\Delta \wp)_{min} =\sqrt{(1+\zeta)\beta_0}.
\end{eqnarray}
The above uncertainty relation can be obtained from the deformed
commutation relation
\begin{eqnarray}\label{gupc}
[\wp,T]=i(1+\beta_0 T^2),
\end{eqnarray}
Because of the extra term $\beta_0 T^2$, this relation cannot be
satisfied by the ordinary price and trend operators since they obey
the canonical commutation relation $[\wp_0,T_0]=i$. However, we can
write them in terms of ordinary operators as
\begin{eqnarray}\label{x0p0}
\wp &=& \wp_0,\\
T &=& T_0 \left( 1 + \frac{1}{3}\beta_0\, T_0^2
\right),\label{x0p01}
\end{eqnarray}
It is easy to check that using this definition, Eq.~(\ref{gupc}) is
satisfied to first-order of GUP parameter i.e.
\begin{eqnarray}
[\wp,T]=i(1+\beta_0 T^2)+{\cal{O}}(\beta_0^2),
\end{eqnarray}
where $\wp$ and $T$ are given by Eqs.~(\ref{x0p0}) and
(\ref{x0p01}). Note that since $[\wp,T]\ne i$, we cannot further
consider the generalized trend operator $T$ as the derivative with
respect to
price\footnote{$T=-i\displaystyle\frac{\partial}{\partial\wp_0}+i\frac{\beta_0}{3}\frac{\partial^3}{\partial\wp_0^3}.$}.
Now using Eqs.~(\ref{gup},\ref{gupc}) and $\Delta \wp\Delta T\ge
(1/2)|\langle[\wp,T]\rangle|$ we find $\zeta=\beta_0\langle
T\rangle^2$. So using Eq.~(\ref{zeta}) the absolutely smallest
uncertainty in price is
\begin{eqnarray}
(\Delta \wp)_{min} =\sqrt{\beta_0},
\end{eqnarray}
when the expectation value of the trend operator (or $\zeta$)
vanishes, namely $\langle T\rangle=0=\zeta$. We can interpret
$(\Delta \wp)_{min}$ as the minimal price length and indicates that
we cannot measure the price of a stock with uncertainty less than
$(\Delta \wp)_{min}$ which agrees with discrete nature of the stock
price. It is worth mentioning that, the generalized uncertainty
relation also appears in the context of quantum gravity where there
is a minimal observable length proportional to the Planck length
\cite{Kempf,ref1,ref2,ref3,ref4}. In the string theory one can
interpret this length as the length of strings.

Since  $\wp$ and $T$ do not exactly satisfy Eq.~(\ref{gupc}), our
approach is essentially perturbative. Obviously, this procedure
affects all Hamiltonians in quantum financial models. To proceed
further, let us consider the following Hamiltonian:
\begin{eqnarray}
H=\frac{T^2}{2m} + V(\wp,t),
\end{eqnarray}
which using Eq.~(\ref{x0p0}) can be written as
\begin{eqnarray}\label{Hgup}
H=H_0+\beta_0 H_1+{\cal{O}}(\beta_0^2),
\end{eqnarray}
where $H_0=\frac{\displaystyle T_0^2}{\displaystyle2m} +
V(\wp_0,t)$ and $H_1=\frac{\displaystyle T_0^4}{\displaystyle3m}$.
So the corrected term in the modified Hamiltonian is only trend
dependent and is proportional to $T_0^4$. In fact, the presence of
this term leads to a positive shift in energy spectrum.

\section{The Schr\"odinger equation}
Many factors impact the price and the trend of the stock in the
financial markets. These factors include political environment,
market information, economic policies of the government, psychology
of traders, etc. So it is difficult to construct a Hamiltonian which
contains all effective variables. However, we can write a simple
Hamiltonian to model the fluctuation of the stock price with ideal
periodic external factors.

In most Chinese stock markets there is a price limit rule: the rate
of return in a trading day compared with the previous day's closing
price cannot be more than $\displaystyle\pm10\%$. So the stock price
fluctuates between the price limits or in a one-dimensional infinite
well (particle in a box). The size of the box is
$d_0=\bar{\wp}\times 20\%$ where $\bar{\wp}$ is the previous day's
closing price. Now if we use a transformation of coordinate
$\wp'=\wp_0-\bar{\wp}$, the infinite square well will be symmetric
with width $d$ and we can define the absolute return as
$r=\wp'/\bar{\wp}$. So the rate of the return is the new coordinate
variable and the well's width becomes $d=20\%$. Now we can write the
GUP corrected Hamiltonian inside the well in the absence of external
factors approximately as
\begin{eqnarray}
\hat{H}=-\frac{1}{2m}\frac{\partial^2}{\partial r^2}+
\frac{\beta}{3m}\frac{\partial^4}{\partial r^4},
\end{eqnarray}
where $m=m_0/\bar{\wp}^2$, $\beta=\beta_0/\bar{\wp}^2$, and it is
valid to first-order of GUP parameter (\ref{Hgup}). This Hamiltonian
has exact eigenvalues and eigenfunctions \cite{ref3}
\begin{eqnarray}
\label{soll1}\phi_n(r)&=&\sqrt{\frac{2}{d}}\sin\left(\frac{n\pi(r+d/2)}{d}\right),\\
\label{soll2}E_n&=&\frac{n^2\pi^2}{2md^2}+\beta\frac{n^4\pi^4}{3md^4},
\end{eqnarray}
where $n=1,2,3,\ldots$. To write the total Hamiltonian we need to
add the potential which describes the effects of information on the
stock price. The market information usually results either in the
increase of the stock price or in the decrease of the stock price.
Here, similar to Ref.~\cite{Zhang}, we consider a periodical
idealized model which represents the two types of information. This
form of potential also appears for a charged particle moving in an
electromagnetic field with the difference that the information play
the role of the external fields. So up to the dipole approximation
we can write the GUP corrected Hamiltonian of this coupled system as
\begin{eqnarray}\label{HE}
\hat{H}=-\frac{1}{2m}\frac{\partial^2}{\partial r^2}+
\frac{\beta}{3m}\frac{\partial^4}{\partial r^4}+\lambda
\,r\cos\omega t,
\end{eqnarray}
where $\omega$ is the frequency of information and $\lambda$ denotes
the amplitude of the information field. The first two terms of the
above equation represent the GUP corrected kinetic energy of the
stock return and the last term corresponds to the potential energy
due to presence of information in the stock market. Note that, the
choice of the Hamiltonian in (\ref{HE}) is not the only one and we
can replace $\cos(\omega t)$ with $\sin(\omega t)$, as well as with
some other periodic functions.

To find the temporal evolution of the wave function in
price-representation, we need to solve the following Schr\"odinger
equation
\begin{eqnarray}
i\frac{\partial}{\partial
t}\psi(r,t)=\left[-\frac{1}{2m}\frac{\partial^2}{\partial r^2}+
\frac{\beta}{3m}\frac{\partial^4}{\partial r^4}+\lambda\,
r\cos\omega t\right]\psi(r,t).
\end{eqnarray}
To solve this equation, we can use the perturbative procedure that
is also used in Ref.~\cite{14} in connection with stock markets.
Since the exact solutions for $\lambda=0$ is presented in
Eqs.~(\ref{soll1}) and (\ref{soll2}), we can expand the solutions in
terms of these state vectors \cite{21}
\begin{eqnarray}
\psi(r,t)=\sum_n c_n(t)e^{-i E_nt}\phi_n(r),
\end{eqnarray}
where
\begin{eqnarray}
c_n(t) = c_n(0) -i \lambda \sum_k \langle n|r|k\rangle \int_0^t \mathrm{d}t'
c_k(t')\cos\omega t'\, e^{-i(E_k - E_n)t'}.
\end{eqnarray}
By repeatedly substituting this expression back into right hand
side, we obtain an iterative solution
\begin{eqnarray}
c_n(t) = c_n^{(0)}+ c_n^{(1)} + c_n^{(2)} + \ldots,
\end{eqnarray}
where, for instance, $c_n^{(0)}=c_n(0)$ and the first-order term is
\begin{eqnarray}
c_n^{(1)}(t) =-i\lambda \sum_k \langle n|r|k\rangle \int_0^t \mathrm{d}t'
c_k(0)\cos\omega t'\, e^{-i(E_k - E_n)t'}.
\end{eqnarray}
If we take the initial wave function as the ground state of
unperturbed Hamiltonian (a cosine distribution to simulate the state
of stock price in equilibrium) i.e. $\psi(r,0)=\langle
r|1\rangle=\displaystyle\sqrt{\frac{2}{d}}\cos\left(\frac{\pi
r}{d}\right)$, we have $c_n(0)=\delta_{1n}$ which results in \cite{14}
\begin{eqnarray}
c_n^{(1)}(t) =-i\lambda  \langle n|r|1\rangle \int_0^t \mathrm{d}t'
\cos\omega t'\, e^{-i(E_1 - E_n)t'},
\end{eqnarray}
where $\langle
n|r|1\rangle=-\displaystyle\frac{8nd}{(n^2-1)^2\pi^2}$ for $n$ even
and $\langle n|r|1\rangle=0$ for $n$ odd. By evaluating the time
integral we get
\begin{eqnarray}
c_n^{(1)}(t) =\lambda\frac{4nd}{(n^2-1)^2\pi^2}
\left(\frac{e^{i\left[E_n-E_1+\omega\right]t}-1}{E_n-E_1+\omega}
+\frac{e^{i\left[E_n-E_1-\omega\right]t}-1}{E_n-E_1-\omega}\right),
\end{eqnarray}
for $n$ even and $c_n^{(1)}(t)=0$ for $n$ odd. As this equation
shows, at the characteristic frequency $\omega_n=E_n-E_1$ we observe
a large transition probability from ground state to $(n-1)$th
excited state, namely
\begin{eqnarray}
\omega_n=\omega_n^0\left(1+\frac{4}{3}\beta
m\frac{n^2+1}{n^2-1}\omega_n^0\right),
\end{eqnarray}
where $\omega_n^0=\displaystyle\frac{\pi^2}{2md^2}(n^2-1)$ are the
characteristic frequencies at the continuous limit. In the Chinese
stock market the average stock price is approximately $10$ Yuan and
$(\Delta \wp)_{min}=0.01$ Yuan. To find $m$ we need to calculate the
mean daily volatility from annual volatility given by
\begin{eqnarray}
\sigma_{\mathrm{daily}}=\sqrt{\frac{1}{252}}\sigma_{\mathrm{annual}},
\end{eqnarray}
since there are $252$ trading days in any given year. For instance,
the annual volatility of Chinese stock market during 2001--2002 was
about $0.3\%$ \cite{Xu} which results in
$m=\sigma_{\mathrm{daily}}^{-2}\simeq3\times10^{3}$. So we obtain
$\beta_0=(\Delta \wp)^2_{min}=100\beta\simeq10^{-4}$ and
$\omega_n^0\simeq4\times10^{-3}(n^2-1)\,\mbox{s}^{-1}$. In the
presence of the minimal trading value we observe a frequency
dependent positive shift in the characteristic frequencies
proportional to $\omega_n^0$ as
$\omega_n\simeq\omega_n^0\left[1+4\times 10^{-3}\left(\frac{n^
2+1}{n^2-1}\right)\,\omega_n^0\right]$ or
\begin{eqnarray}
\omega_n\simeq\omega_n^0\left(1+4\times
10^{-3}\omega_n^0\right),
\end{eqnarray}
for large $n$. Since $\omega_n>4\times10^{-3}\,\mathrm{s}^{-1}$ if a
single cycle of information fluctuation is larger than $25$ minutes
there is no large transition probability to other states and the
probability density of the rate of stock return approximately
maintains its shape over time (see Fig.~2 of Ref.~\cite{Zhang} for
$\omega=10^{-4}\,\mathrm{s}^{-1}$).

Note that, in quantum gravity, it is usually assumed that the
minimal length is of the order of the Planck length $\ell_{Pl}\sim
10^{-35}\,m$. However, the existence of this infinitesimal length is
not yet confirmed by the experiment. On the other hand, in quantum
finance the minimal trading value is not too small which makes
essentially detectable effects. In other words, the application of
GUP in quantum description of finance is more meaningful than in
quantum physics.

\section{Conclusions}
We have studied the effects of the discreteness of the stock price
on the quantum models for the stock markets. In this formalism, the
minimum trading value of every stock is not zero and the stock price
and its trend satisfy the generalized uncertainty relation. This
modifies all Hamiltonians of the stock markets and adds a term
proportional to the fourth power of the trend to the Hamiltonians.
For the quantum model proposed by Zhang and Huang where there is a
price limit rule and the information has a periodic fluctuation, we
obtained the characteristic frequencies of the quantum system. If
the frequency of information fluctuation coincides with $\omega_n$
we have a large transition probability to $(n-1)$th excited state.
We also showed that the discrete nature of the stock price results
in a positive frequency dependent shift in characteristic
frequencies where for the Chinese stock market we have
$\omega_n>4\times10^{-3}\,\mathrm{s}^{-1}$.


\begin{thebibliography}{99}
\bibitem{1}     R.N. Mantegna and H.E. Stanley, An Introduction to Econophysics: Correlations and Complexity in Finance, Cambridge University Press, Cambridge, 1999.
\bibitem{4}     K. Ilinski, Physics of Finance, Wiley, New York, 2001.
\bibitem{5}     B.E. Baaquie, Quantum Finance, Cambridge University Press, Cambridge, 2004.
\bibitem{9}     C. Ye and J.P. Huang, Non-classical oscillator model for persistent fluctuations in stock markets, Physica A 387 (2008) 1255–-1263.
\bibitem{10}    A. Ataullah, I. Davidson, and M. Tippett, A wave function for stock market returns, Physica A 388 (2009) 455–-461.
\bibitem{11}    F. Bagarello, Stock markets and quantum dynamics: a second quantized description, Physica A 386 (2007) 283-–302.
\bibitem{12}    F. Bagarello, An operatorial approach to stock markets, J. Phys. A 39 (2006) 6823–-6840.
\bibitem{13}    F. Bagarello, The Heisenberg picture in the analysis of stock markets and in other sociological contexts, Qual. Quant. 41 (2007) 533–-544.
\bibitem{14}    F. Bagarello, A quantum statistical approach to simplified stock markets, Physica A 388 (2009) 4397–-4406.
\bibitem{15}    F. Bagarello, Simplified stock markets described by number operators, Rep. Math. Phys. 63 (2009) 381–-398.
\bibitem{6}     M. Schaden, Quantum finance, Physica A 316 (2002) 511–-538.
\bibitem{7}     D. Meyer, Quantum strategies, Phys. Rev. Lett. 82 (1999) 1052--1055.
\bibitem{8}     J. Eisert, M. Wilkens, and M. Lewenstein, Quantum games and quantum strategies, Phys. Rev. Lett. 83 (1999) 3077--3080.
\bibitem{Black} F. Black and M. Scholes, The pricing of options and corporate liabilities, Journal of Political Economy, 81 (1973) 637--654.
\bibitem{Merton}R.C. Merton, Continuous Time Finance, Blackwell, 1990.
\bibitem{Haven} E. Haven, A discussion on embedding the Black-Scholes option pricing model in a quantum physics setting, Physica A 304 (2002) 507--524.
\bibitem{Schaden}M. Schaden, review of the book ``Interest Rates and Coupon Bonds in Quantum Finance'' by B.E. Baaquie, Am. J. Phys. 78 (2010) 654--656.
\bibitem{Zhang} C. Zhang and L. Huang, A quantum model for the stock market, Physica A 389 (2010) 5769–-5775.
\bibitem{Kempf} A. Kempf, G. Mangano, and R.B. Mann, Hilbert space representation of the minimal length uncertainty relation, Phys. Rev. D 52 (1995) 1108--1118.
\bibitem{ref1}  D. Amati, M. Ciafaloni, and G. Veneziano, Can spacetime be probed below the string size?, Phys. Lett. B 216 (1989) 41--47.
\bibitem{ref2}  S. Das and E.C. Vagenas, Universality of quantum gravity corrections, Phys. Rev. Lett. 101 (2008) 221301 [4 pages], arXiv:0810.5333.
\bibitem{ref3}  P. Pedram, A class of gup solutions in deformed quantum mechanics, Int. J. Mod. Phys. D 19 (2010) 2003--2009, arXiv:1103.3805.
\bibitem{ref4}  P. Pedram, On the modification of the Hamiltonians' spectrum in gravitational quantum mechanics, Erouphys. Lett. 89 (2010) 50008 [5 pages], arXiv:1003.2769.
\bibitem{21}    D.J. Griffiths, Introduction to Quantum Mechanics, Prentice Hall, Upper Saddle River, NJ, 1995.
\bibitem{Xu}    Y. Xu, Diversification in the Chinese stock market, Working Paper, School of Management, The University of Texas at Dallas and Shanghai Stock Exchange (2003).
\end{thebibliography}
\end{document}